\documentclass[%
 aip,
rsi,amsmath,amssymb,citeautoscript,reprint,
]{revtex4-1}
\usepackage{graphicx}
\usepackage{dcolumn}
\usepackage{amsmath}
\usepackage{xcolor}
\usepackage{bm}
\usepackage[colorlinks=true,bookmarks=false,citecolor=blue,linkcolor=blue,hyperfootnotes=true,urlcolor=blue]{hyperref}

\begin{document}

\preprint{AIP/123-QED}

\title{Piezoelectric effect and polarization switching in Al$_{1-x}$Sc$_{x}$N}
\author{Haochen Wang}
\affiliation{
Materials Department, University of California, Santa Barbara, California 93106, USA}
\affiliation{
School of Physical Sciences, University of Science and Technology of China, Hefei, Anhui 230026, China}
\author{Nicholas Adamski}
\author{Sai Mu}
\author{Chris G. Van de Walle}
\email{vandewalle@mrl.ucsb.edu}
\affiliation{
Materials Department, University of California, Santa Barbara, California 93106, USA}

\date{\today}
\begin{abstract}
Aluminum nitride is piezoelectric and exhibits spontaneous polarization along the $c$-axis, but the polarization cannot be switched by applying an electric field.
Adding Sc to AlN enhances the piezoelectric properties, and can make the alloy ferroelectric.
We perform a detailed first-principles analysis of spontaneous and piezoelectric polarization.
Comparisons between explicit supercell calculations show that the virtual crystal approximation produces accurate results for polarization, but falls short in describing the phase stability of the alloy.
We relate the behavior of the piezoelectric constant $e_{33}$ to the microscopic behavior of the internal displacement parameter $u$, finding that the internal strain contribution dominates in the Sc-induced enhancement.
The value of $u$ increases with scandium concentration, bringing the alloy locally closer to a layered hexagonal structure.
Our approach allows us to calculate the ferroelectric switching barrier, which we analyze as a function of Sc concentration and temperature based on Ginzburg-Landau theory.

\end{abstract}

\maketitle

\section{Introduction}

The low symmetry of the wurtzite (wz) structure allows AlN to be piezoelectric and also to exhibit spontaneous polarization.
The spontaneous polarization of AlN has a specific orientation along the $c$ axis, and cannot be switched.
Switchable polarization would make the material ferroelectric, but in AlN this cannot be accomplished with an applied electric field below its dielectric breakdown limit.
Adding Sc to AlN qualitatively changes the properties.
It had already been found that ScAlN alloys exhibit enhanced piezoelectricity\cite{akiyama2009enhancement}, with a piezoelectric constant $e_{33}$ larger than piezoelectric materials that are presently in use \cite{tasnadi2010origin}.
Excitingly, Fichtner \textit{et al.}\cite{fichtner2019AlScN} demonstrated ferroelectric switching and reported a high paraelectric transition temperature of 600$^\circ$C for Al$_{0.64}$Sc$_{0.36}$N.
Very recently, the electric-field-induced formation of Al-polar domains in originally N-polar ScAlN thin films has been directly observed using scanning transmission electron microscopy \cite{wolff2021atomic}, providing direct evidence for ferroelectric switching in ScAlN alloys at the atomic scale.

\begin{figure}
    \centering
    \includegraphics[scale=0.34]{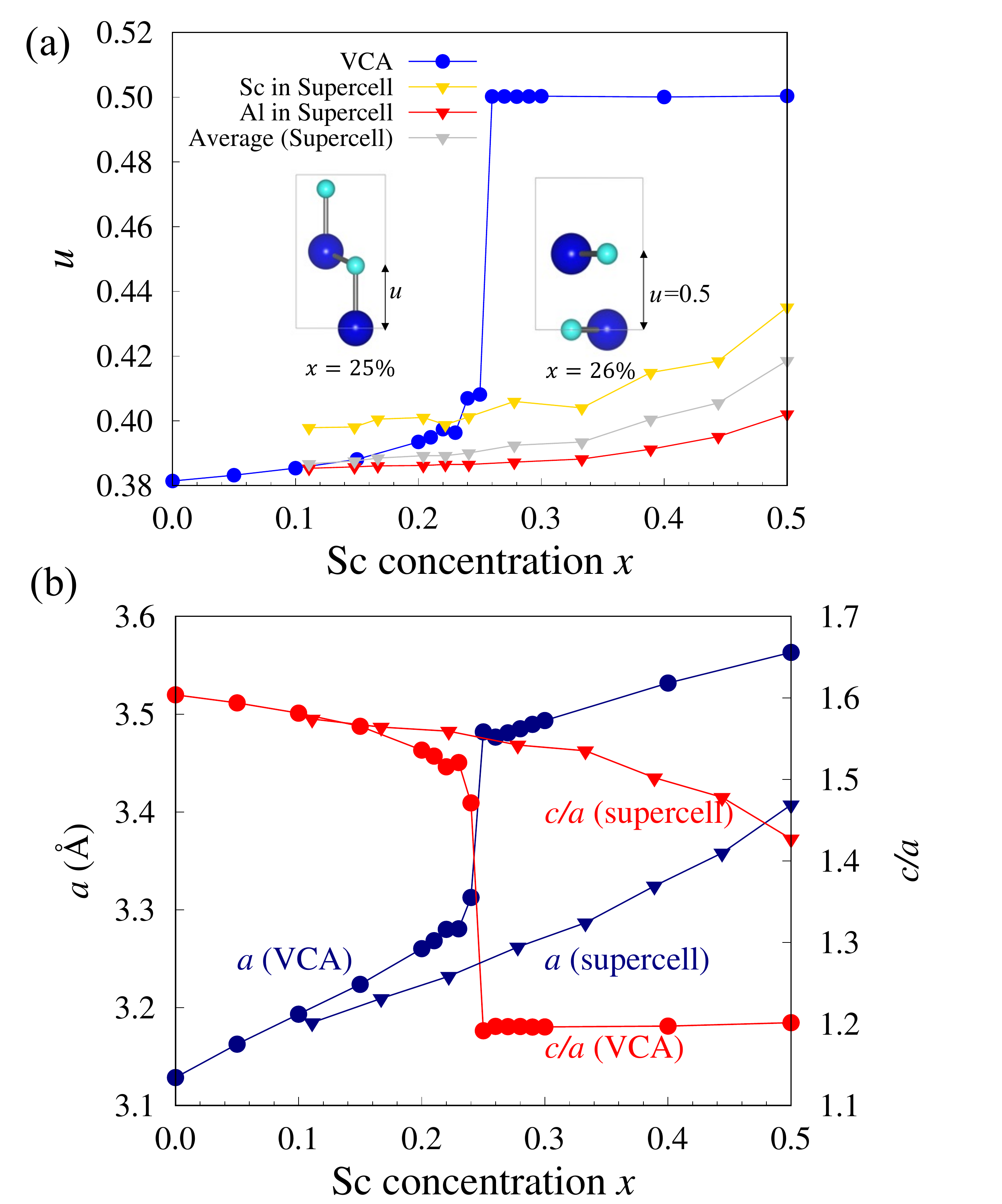}
    \caption{{(a)} Internal displacement parameter $u$ and {(b)} lattice parameter $a$ and $c/a$ ratio as a function of Sc concentration $x$, as obtained using VCA
    and from supercell calculations.
In (a), $u$ values from supercell calculations are shown for individual cations, along with the weighted average.
    The inserts define $u$ and depict the structures of Al$_{1-x}$Sc$_{x}$N near the phase transition point: $x=0.25$ corresponds to the wurtzite phase, $x=0.26$ to the layered hexagonal phase.
}
    \label{u-Sc}
\end{figure}

Unlike AlN, ScN stabilizes in the rocksalt structure, however, experimentally\cite{akiyama2009enhancement,fichtner2019AlScN} it has been found that Al$_{1-x}$Sc$_x$N alloys can maintain the wz structure up to about $x$=0.43.
To analyze the behavior of an alloy, the zeroth order approximation is to interpolate properties between the end points; for Al$_{1-x}$Sc$_x$N, this would mean interpolating between AlN and ScN, both in the wz structure.
Interestingly, ScN cannot be stabilized in the wz structure (space group $P6_3mc$); calculations\cite{farrer2002properties} show that, when initialized in the wz structure, ScN spontaneously relaxes to a layered hexagonal (h) structure ($P6_3/mmc$) in which the anions and cations lie in the same plane.  This can be characterized by the internal displacement parameter $u$ [illustrated in the insets of Fig.~\ref{u-Sc}(a)]: $u$ is the ratio of the bond length along the $c$ axis to the $c$ lattice parameter; $u$ is close to 0.375 in the wz structure and equal to 0.5 in the layered hexagonal structure.

We note that the layered hexagonal structure has inversion symmetry and hence cannot exhibit polarization\cite{biswas2019development}; this renders the enhancement in piezoelectricity upon addition of Sc all the more intriguing.  However, the expected increase in $u$, based on interpolation between wz-AlN and h-ScN, renders ferroelectric switching more feasible.  Switching indeed involves moving anions and cations past each other (e.g., from $u$=0.375 to $u$=0.625), with the midpoint being the $u$=0.5 structure.  The closer the $u$ value of the alloy is to 0.5, the lower the switching barrier will be.

In this paper we report first-principles calculations aimed at elucidating the microscopic mechanisms underlying the enhancement of the piezoelectric response and the potential for ferroelectric switching in Al$_{1-x}$Sc$_x$N alloys.
The calculations are based on density functional theory (DFT), and alloys are treated in two distinct ways: by performing explicit supercell calculations, and by using the virtual crystal approximation (VCA).
We will demonstrate that the combination and comparison of these approaches provides valuable insight into the underlying physics.
We have investigated the phase stability, and the variation of the $u$ parameter as a function of Sc concentration $x$.
We find that, as the Sc concentration increases, the spontaneous polarization of the alloy decreases, and so does the energy barrier for switching the polarization direction.
Our results show that the enhancement of the piezoelectric constant $e_{33}$ of the alloy, which can be attributed to the large Born effective charge $Z^*_{33}$ and the sensitivity of the internal parameter $u$ to external strain,\cite{tasnadi2010origin} is reliably described using VCA.  
The ability to use VCA to describe polarization in alloys offers the prospect of performing large-scale simulations that would otherwise be prohibitively expensive, for instance to model domain-wall motion.

\section{Computational details}
\label{sec:comp}

The DFT calculations were performed using projector augmented wave (PAW) potentials \cite{Blochl1994projector,Kresse1999PAW}, as implemented in the Vienna \textit{Ab-initio} Simulation Package ({\sc vasp}) \cite{kresse1996efficient}, with a plane-wave energy cutoff of 500 eV.
The generalized gradient approximation (GGA) by Perdew, Burke, and Ernzerhof (PBE) \cite{perdew1996GGA} was employed to account for exchange and correlation.
Berry phase calculations\cite{resta1994macroscopic,vanderbilt2000berry-phase} and density functional purturbation theory were used to compute polarization, Born effective charges and piezoelectric tensors.

To investigate alloying effects we employ both the supercell method and the VCA \cite{vanderbilt2000VCA}.
For the supercell approach, we used a 108-atom cell, which is a $3 \times 3 \times 3$ multiple of the four-atom wz unit cell.
We consider both supercells in which Sc and Al atoms are randomly placed on cation sites, as well as supercells based on special quasi-random structures (SQS), which minimize pairwise short-range order up to fourth-nearest cation-cation pairs\cite{zunger1990special,chen2011influence}.
The calculated physical observables were configurationally averaged over different supercells.

In the VCA method, virtual cations interpolate between Al and Sc elements, thus mimicking arbitrary Sc concentrations in Al$_{1-x}$Sc$_{x}$N solid solutions.
The benefit of VCA lies in its computational efficiency, since calculations can be performed in a 4-atom primitive cell without explicit configurational averaging.
It has been demonstrated that the VCA approach can be integrated with the Berry phase calculations, reproducing reliable Born effective charges and piezoelectric coefficients. \cite{vanderbilt2000VCA}

Hellmann-Feynman forces on each atom were converged to 10 meV/{\AA}.
Brillouin-zone integration for the primitive cell were performed using a 8$\times$8$\times$5 k-point mesh; for the supercells the $\Gamma$ point was used as a single $k$ point.

\section{\label{Results and discussions}Results and discussions}
\subsection{\label{sec:u} Structural parameters}

In Fig.~\ref{u-Sc} we plot the internal displacement parameter $u$, lattice parameter $a$ and the $c/a$ ratio as a function of Sc concentration.
The value of the internal displacement parameter $u$ is 0.381 in wz-AlN and 0.500 in layered hexagonal ScN.
When Sc is added to wz-AlN, $u$ will increase.
We studied the evolution of $u$ by gradually increasing the Sc concentration and determining the most stable structure for each concentration.
As we will discuss in Sec.~\ref{sec:eprofile}, in the VCA calculations the wz structure can be stabilized up to a point between 25\% and 26\% Sc.
We note that our calculations do not address a possible phase transition to the rocksalt structure.  At higher Sc concentrations, rocksalt should be most stable.  However, since we start with the wz structure, there are not enough degrees of freedom during the structural optimization to transition to a rocksalt structure. Still, our calculations address the regime that is most relevant for exploring the polarization properties.

We first analyze the results obtained using the VCA.
Twenty-one different Sc concentrations were calculated.
$u$ and $a$ increases monotonically with Sc concentration $x$,
up to a point between 25\% and 26\% Sc where a discontinuity occurs and $u$ jumps to 0.5.
At 25\% the crystal structure can still be stabilized in a wz phase.
At 26\% the system can no longer be stabilized in a wz phase, and the structure spontaneously relaxes to a hexagonal phase with $u$=0.5.
This layered hexagonal phase ($P6_3/mmc$) has a larger $a$ and a smaller $c/a$ value than wz.
Our VCA result for the concentration at which this transition occurs is lower than the reported experimental value of $\sim$43\%\cite{akiyama2009enhancement,fichtner2019AlScN} and from our calculations using supercells, which indicate the transition occurs near 50\%$\sim$60\%; the latter result is consistent with the calculations of Zhang \emph{et al.}~\cite{zhang2013tunable} who identified the transition at 56\%.
A similar shortcoming of VCA in predicting relative stability of different phases was observed in calculations of Pb(Zr$_{0.5}$Ti$_{0.5}$)O$_3$ alloys.\cite{ramer2000application,ramer2000virtual}

Figure~\ref{u-Sc} also shows results
obtained from supercell calculations.
At low Sc concentrations, the supercell values agree reasonably well with the VCA results.
However, deviations occur when $x$ approaches 25\%, related to the lack of stability of the wurtzite phase in VCA calculations when $x >$25\%.

\subsection{\label{sec:eprofile}Energy profiles}

\begin{figure}[]
    \centering
    \includegraphics[scale=0.6]{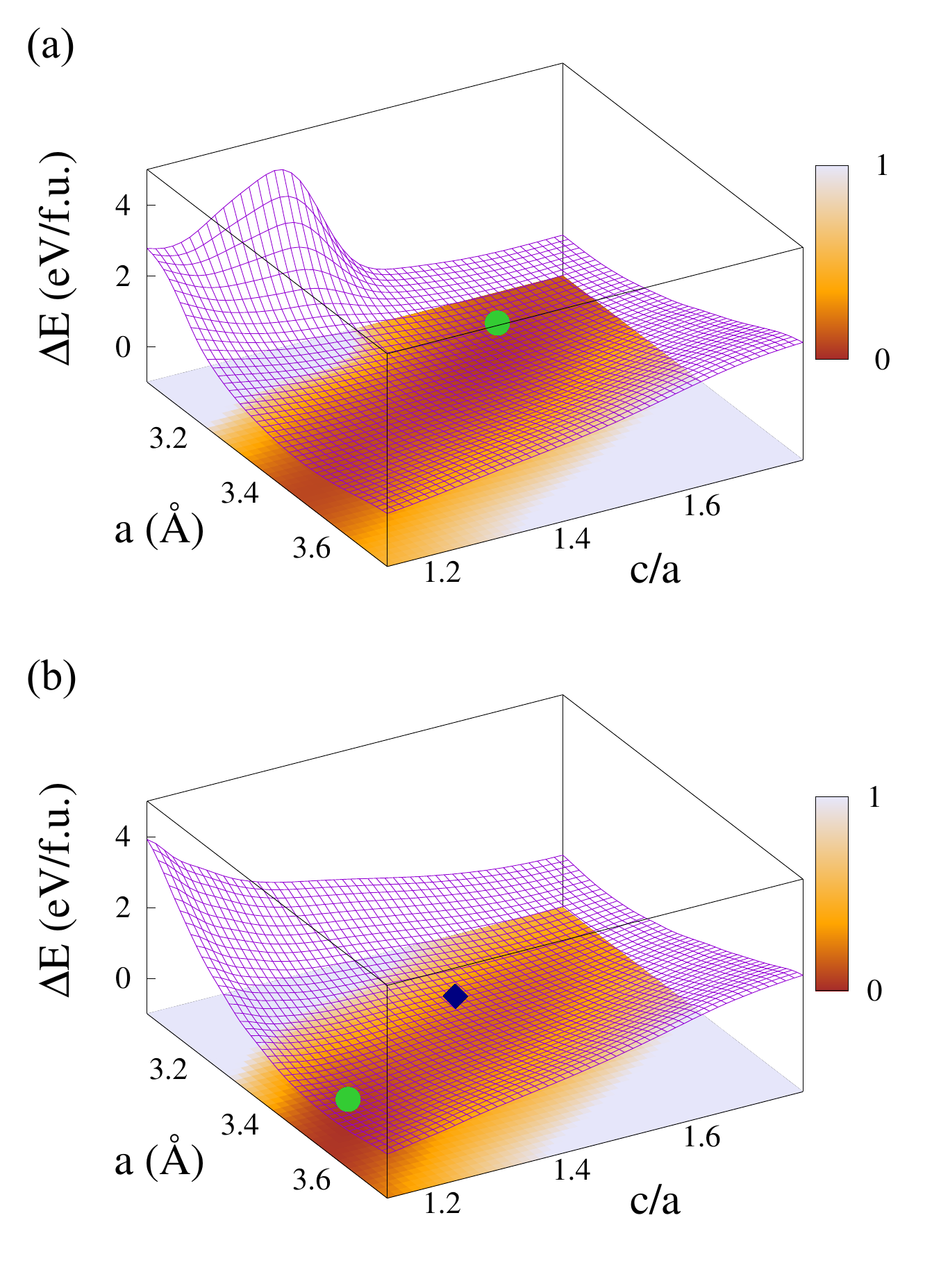}
    \caption{Energy profiles [in eV per formula unit (f.u.)] from VCA calculations for Al$_{1-x}$Sc$_{x}$N with (a) $x=10\%$ and (b) $x=25\%$.
    Green circles denote the global minimum while the blue diamonds denote metastable states.
    }
    \label{energy profiles}
\end{figure}

Using VCA, we calculated the energy profiles of Al$_{1-x}$Sc$_x$N alloys for Sc concentrations ranging from 5\% to 25\%.
Figure~\ref{energy profiles} illustrates the energy profile as a function of the lattice parameters $a$ and $c/a$ for $x=10\%$ [Fig.~\ref{energy profiles}(a)] and $x=25\%$ [Fig.~\ref{energy profiles}(b)].
We note that the calculated energy profiles in both cases are flat over a wide range of lattice parameters.
The flattening of the energy landscape due to a competition between the wz and the hexagonal phase contributes to the enhancement of the piezoelectric coefficient in Al$_{1-x}$Sc$_x$N alloys\cite{tasnadi2010origin}.

For 10\% Sc, the global minimum is found to be a wz structure with internal displacement parameter $u=0.385$ [see Fig.~\ref{energy profiles}(a)].
For $ x \geq 10\%$, the hexagonal structure becomes the ground state, but a wz structure can be stabilized as long as $ x \le 25\%$ [see Fig.~\ref{energy profiles}(b)].
For $x = 25\%$, the global minimum corresponds to a hexagonal phase with $u=0.500$, but we can also stabilize a metastable state, which has the wz phase with $u=0.408$.
When $ x > 25\%$, the wz structure can no longer be stabilized.
Figures~\ref{u-Sc} and \ref{energy profiles} show that the hexagonal structure has a larger $a$ lattice parameter and a smaller $c/a$ ratio than the wz structure.

\subsection{Spontaneous polarization}
\label{sec:sppol}

\begin{figure}
    \centering
    \includegraphics[scale=0.35]{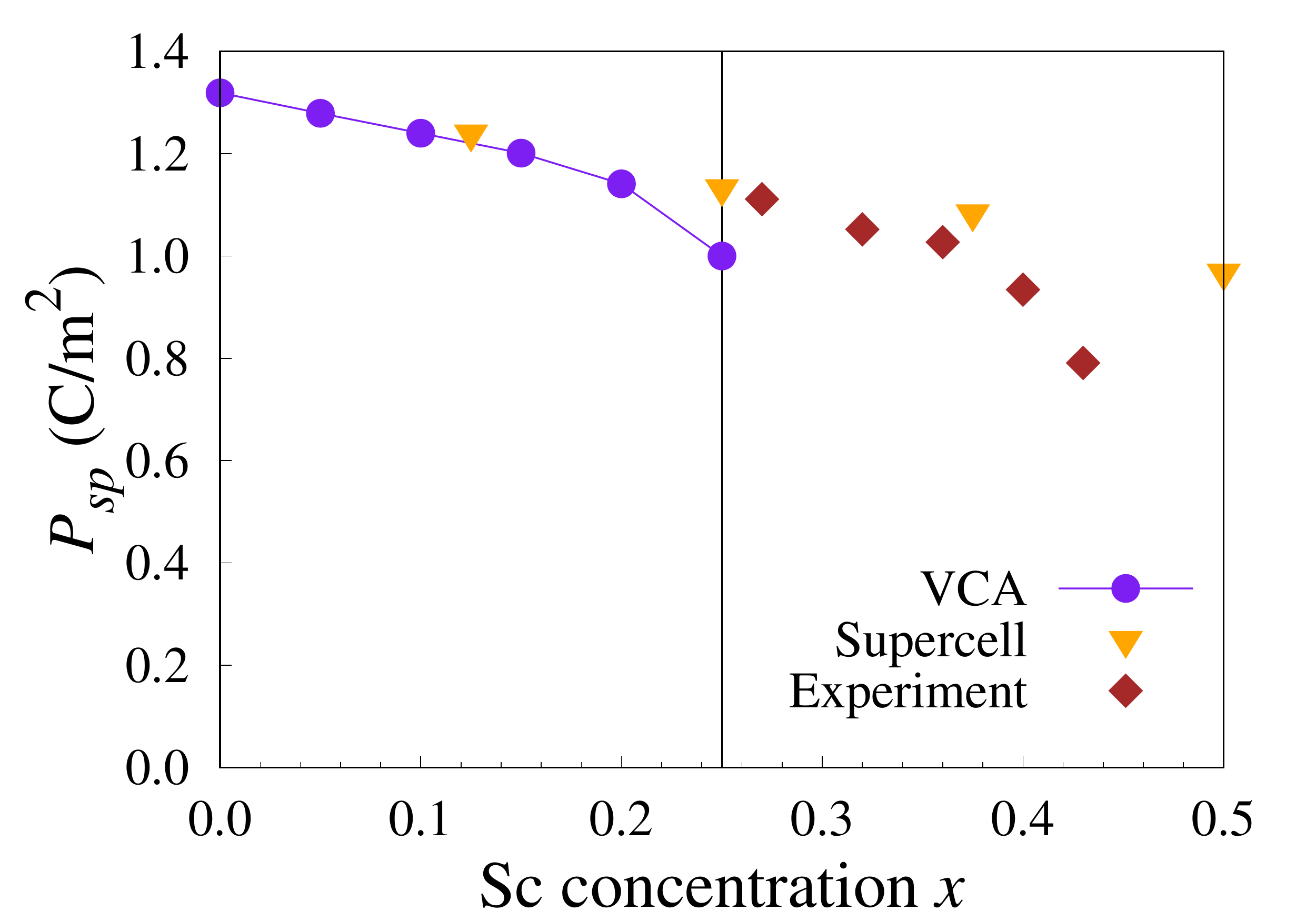}
    \caption{Spontaneous polarization $P_{sp}$ of Al$_{1-x}$Sc$_x$N as a function of Sc concentration $x$.
    Purple circles represent VCA results while triangles are supercell results.  The data point at $x=0$ was calculated for pure AlN. Experimental data points are extracted from the work of Fichtner \textit{et al.}\cite{fichtner2019AlScN} and represented as brown diamonds.}
    \label{polarization}
\end{figure}

Figure~\ref{polarization} shows the results of our calculations of spontaneous polarization $P_{sp}$ in Al$_{1-x}$Sc$_x$N alloys with concentrations of Sc up to 25\%, i.e., the range over which the wz phase is stable (or metastable).
We first examined the polarization of pure AlN and found the spontaneous polarization $P_{sp}$ (expressed per unit cell area) to be extremely close to the result of Ref.~\onlinecite{dreyer2016correct}, which was calculated using the hybrid functional of Heyd, Scuseria and Ernzerhof \cite{HSE}.
At each concentration $x$, we followed the approach of Ref.~\onlinecite{dreyer2016correct}: the structure was optimized; the reference structure for calculating $P_{sp}$ was chosen to correspond to the layered hexagonal structure (which has zero polarization) at the same lattice parameters; and to ensure we stay on a specific branch of the multivalued polarization, we constructed an array of interpolated structures with fixed lattice parameters but different internal displacement parameters $u$.
The resulting values of energy and polarization as a function of $u$ will also serve to map out the switching barrier, as described in Sec.~\ref{sec:barrier}.

Over the range from 0 to 25\% Sc, the polarization decreases monotonically with increasing scandium concentration.
Figure~\ref{polarization} displays results obtained with VCA and with supercells.
Up to $x$=0.2, the VCA results are approximately linear as a function of $x$.
For $x=0.125$, the agreement between VCA and supercell results is very good, but a deviation is evident at 25\%.
The supercell result seems to follow the roughly linear decrease with $x$, while the VCA result is much lower.
The deviation in the VCA result is related to the lack of stability of the wz structure in VCA, as discussed in Sec.~\ref{sec:u}; right above $x$=0.25, the layered hexagonal structure, which has zero polarization, becomes more favorable.

\subsection{Piezoelectric polarization}
\label{sec:pzpol}

The piezoelectric constants $e_{ij}$ can be decomposed into a clamped-ion term and internal-strain contributions \cite{resta1994piezoelectricity,momida2016strong}:
\begin{equation}
    \begin{split}
        e_{ij} &= e_{ij}^{\rm clamped} + e_{ij}^{\rm int}\\
        &= \left.\frac{\partial P_{i}}{\partial \varepsilon_{j}}\right|_{u}+\left.\sum_{\alpha, k} \frac{\partial P_{i}}{\partial u_{\alpha, k}}\right|_{\varepsilon} \frac{\partial u_{\alpha, k}}{\partial \varepsilon_{j}} \, .
        \label{eq:eij}
    \end{split}
\end{equation}
$P_{i}$ is the polarization along the $i$ direction and $\varepsilon_{j}$ is the strain along the $j$ direction.
$u_{\alpha, k}$ is the internal atomic coordinate along the $k$ direction for atom $\alpha$ in the cell. Note that $u_{\alpha, k}$ is distinct from the wurtzite internal displacement parameter $u$ as defined in Fig.~\ref{u-Sc}, but they are related by, e.g.,
\begin{equation}
    u_{\mathrm{Sc}}=u_{\mathrm{Sc}, 3}-u_{\mathrm{N}, 3} \, .
        \label{eq:defu}
\end{equation}
The clamped-ion term arises from the contributions of electrons when the ions are frozen at their zero-strain equilibrium internal atomic coordinates $u_{\alpha,k}$, i.e., the ionic coordinates follow the strain-induced deformation of the lattice vectors.
The internal-strain contributions term reflects the distortion of the ionic coordinates at fixed strain.
We note that the first factor in the internal-strain term can be rewritten in terms of the Born effective charge $Z^*$, which is defined as
\begin{equation}
    Z_{\alpha, i j}^{*}=\frac{V}{q_{e}} \frac{\partial P_{i}}{\partial r_{\alpha, j}} \, .
\label{eq:Z}
\end{equation}
$r_{\alpha, j}$ is the Cartesian atomic coordinate along the $j$ direction of atom $\alpha$, $V$ is the volume of the unit cell, and $q_e$ is the electron charge.

Here we focus on piezoelectric polarization along the $z$ axis and hence examine $e_{33}$:
\begin{equation}
\begin{split}
    e_{33}=\left.\frac{\partial P_{3}}{\partial \varepsilon_{3}}\right|_{u}+\sum_{\alpha} \frac{q_{e} c Z_{\alpha, 33}^{*}}{V} \frac{\partial u_{\alpha, 3}}{\partial \varepsilon_{3}}  \\
    = \left.\frac{\partial P_{3}}{\partial \varepsilon_{3}}\right|_{u}+ \frac{4q_{e} Z_{33}^{*}}{\sqrt{3} a^2} \frac{\partial u}{\partial \varepsilon_{3}}  \, .
        \label{eq:e33}
\end{split}
\end{equation}
The transition from internal atomic coordinates $u_{\alpha, 3}$ to Cartesian coordinates $r_{\alpha,3}$ in the definition of $Z^*$ has introduced the factor $c$ (lattice parameter).
In the second line of Eq.~(\ref{eq:e33}) the internal-strain contribution is rewritten in terms of the wurtzite internal displacement parameter $u$ and $Z^*_{33}$, the Born effective charge for the virtual cation. The factor $a^2$ in the denominator arises from expressing the volume of the wz unit cell $V$ in terms of the lattice parameters.

\begin{figure}
    \centering
    \includegraphics[scale=0.9]{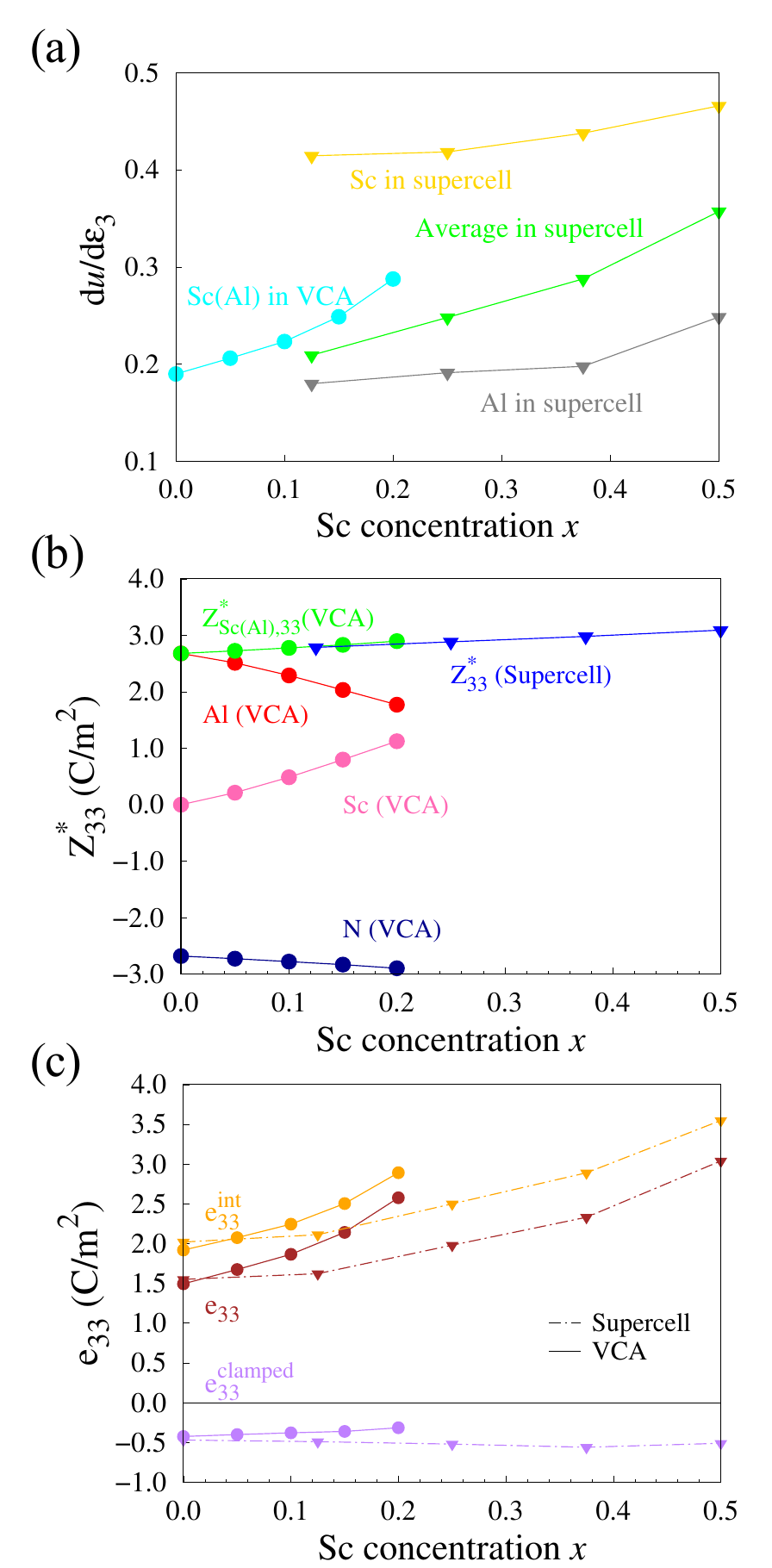}
    \caption{Contributions to the piezoelectric coefficient $e_{33}$ as a function of Sc concentration calculated with VCA and with supercells; supercell results are adapted from Ref.~\onlinecite{tasnadi2010origin}.
    (a) Response of the wurtzite internal displacement parameter $u_{\alpha}$ for species $\alpha$ [as defined in Eq.~(\ref{eq:defu})] to strain $\varepsilon_3$.  VCA results are shown in light blue, and results of supercells are distinguished by using yellow triangles for Sc, grey triangles for Al and green triangles for averaged cation.
    (b) Born effective charge of cations as a function of Sc concentration obtained using VCA (colored circles) and the supercell method (blue triangles).
    (c) Different contributions to $e_{33}$: clamped-ion (purple), internal strain (orange), and total (brown).
    The VCA and supercell results are distinguished using solid and dashed lines.
    }
    \label{piezo}
\end{figure}

Figure~\ref{piezo} displays our results for $e_{33}$ and its various contributions calculated using VCA.
We compare with results obtained using supercells in Ref.~\onlinecite{tasnadi2010origin}.
$du/d\varepsilon_3$ is shown in Fig.~\ref{piezo}(a).  The VCA results show a striking dependence on $x$: $du/d\varepsilon_3$ increases markedly with $x$.
After averaging over the Sc and Al values, the supercell results (green triangles) are close to the VCA results in the low-$x$ limit but again deviate as $x$ increases towards 0.25.

The much larger values of $du/d\varepsilon_3$ for Sc in the supercells can be explained as follows:
We know that ScN prefers the (hexagonal) $u=0.5$ structure, in which the Sc and N atoms are threefold coordinated, and in which the $a$ lattice parameter is significantly larger than in the AlN wz structure.
In Al$_{1-x}$Sc$_x$N alloys with low Sc concentrations, $a$ will be close to the value in wz AlN, and hence the position of the Sc atom is constrained; the corresponding $u$ value will be farther from 0.5 than Sc prefers.  When strain is applied, the atoms will eagerly take advantage of the distortion to locally adopt a structure closer to their preferred environment, leading to a large and positive $du/d\varepsilon$ value.
For Al atoms, $u$ is closer to its preferred value, and hence Al atoms do not experience such a strong driving force to change their $u$ parameter, explaining the smaller magnitude of the $du/d\varepsilon$ values.
In spite of the distinct structural mechanisms experienced by Al and Sc atoms, the VCA does a decent job reproducing the weighted average of $du/d\varepsilon_3$ for the cations in the supercells.  The deviation as $x$ approaches 25\% is again due to the exaggerated stability of the hexagonal phase in VCA.

For the Born effective charge $Z^*_{33}$ [Fig.~\ref{piezo}(b)], we see that the dependence on Sc concentration is roughly linear.
We found the average value of the cation Born effective charges calculated in VCA (green dots) to be in excellent agreement with the supercell results (blue triangles). This consistency highlights the power and efficiency of the VCA in accurate prediction of the Born effective charges.

$Z_{33}^*$ and $du/d\varepsilon$ are combined to calculate $e_{33}^{\rm int}$, which is plotted as a function of $x$ in Fig.~\ref{piezo}(c), along with
$e_{33}^{\rm clamped}$ as well as the total $e_{33}$.
The clamped-ion term $e_{33}^{\rm clamped}$ is negative and significantly smaller in magnitude than $e_{33}^{\rm int}$.
The supercell results for $e_{33}^{\rm clamped}$ show virtually no variation with $x$.
The VCA results are reasonably close to the supercell results, but show a slight decrease in the magnitude of $e_{33}^{\rm clamped}$ as $x$ increases towards 0.25, the concentration at which (within VCA) the polarization disappears.
$e_{33}^{\rm int}$ is both larger in magnitude and varies more quickly with $x$ than $e_{33}^{\rm clamped}$.
The increase in $e_{33}^{\rm int}$ is overestimated in VCA, as compared to supercell results.  This can be directly attributed to the overestimate of $du/d\varepsilon$ in VCA [Fig.~\ref{piezo}(a)] discussed above.

\begin{figure*}
    \centering
    \includegraphics[scale=0.5]{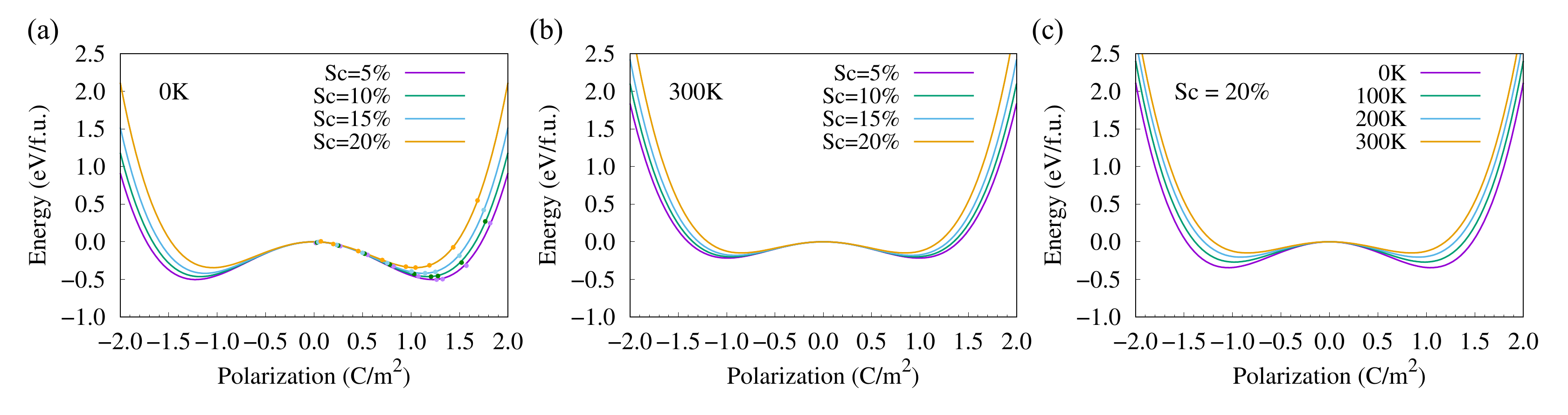}
    \caption{
    (a) Energy per formula unit (f.u.) of Al$_{1-x}$Sc$_{x}$N at $T$=0 as a function of polarization for various Sc concentrations.  The data points (shown only for positive polarization) at each Sc concentration are based on VCA calculations as described in the text.  The curves are fits to Eq.~(\ref{landau}).
    (b) Energy per f.u. as a function of polarization predicted by phenomenological modeling according to Eq.~(\ref{landau}) at $T$=300 K, for different Sc concentrations.
    (c) Free energy per f.u. as a function of polarization according to Eq.~(\ref{landau}) for Al$_{0.8}$Sc$_{0.2}$N, for different temperatures.
    In all plots the energy at the point of zero polarization has been set to zero.
    }
    \label{energy barriers}
\end{figure*}

\subsection{Ferroelectric switching barrier}
\label{sec:barrier}

For a wz structure, energy minima occur for values of the internal displacement parameters $u$ and $1-u$; these represent degenerate and symmetric states with opposite polarization, corresponding to the minima of a double-well potential.
Our calculated double-well potentials as a function of polarization (proportional to $u$) for different Sc concentration are shown in Fig.~\ref{energy barriers}(a).
As each Sc concentration, we freeze the lattice parameters $a$ and $c$ and perform VCA calculations for total energy and polarization as a function of $u$.
The symmetric structure at $u$=0.5, which has zero polarization, constitutes the barrier;
the well depth thus corresponds to the energy barrier to switch the polarization.

The switching path we are considering here is restricted in the sense that all the dipoles in the material are switched simultaneously and uniformly.
In real systems, ferroelectric switching usually involves initial nucleation of a reversed ferroelectric domain followed by gradual switching of the global polarization via domain-wall motion.
The switching barrier for this non-uniform process is usually much lower than the barrier calculated here.
In addition, our calculations are performed assuming the lattice parameters $a$ and $c$ are kept fixed during the switching.
This lack of structural flexibility will yield a higher switching barrier.
Therefore, our calculated barrier represents an upper bound on the actual barrier.
However, our calculations provide insight into how the Sc concentration affects the ability to switch polarization.

Further insight into the switching process at finite temperatures can be obtained by employing the Ginzburg-Landau theory for ferroelectric-paraelectric phase transitions \cite{landau1936phase}.  The free energy of the system $E(T,P)$ as a function of polarization is expressed as:
\begin{equation}\label{landau}
    E(T, P)=E_{0}(T)+\frac{1}{2} a(T) P^{2}+\frac{1}{4} b(T) P^{4}+\cdots \, .
\end{equation}
$P$ is the polarization along $c$ direction, and $E_0(T)$ is the free energy of system with zero polarization.
$a(T)$ and $b(T)$ are the coefficients of the second and fourth order terms.

\begin{figure}
    \centering
    \includegraphics[scale=0.32]{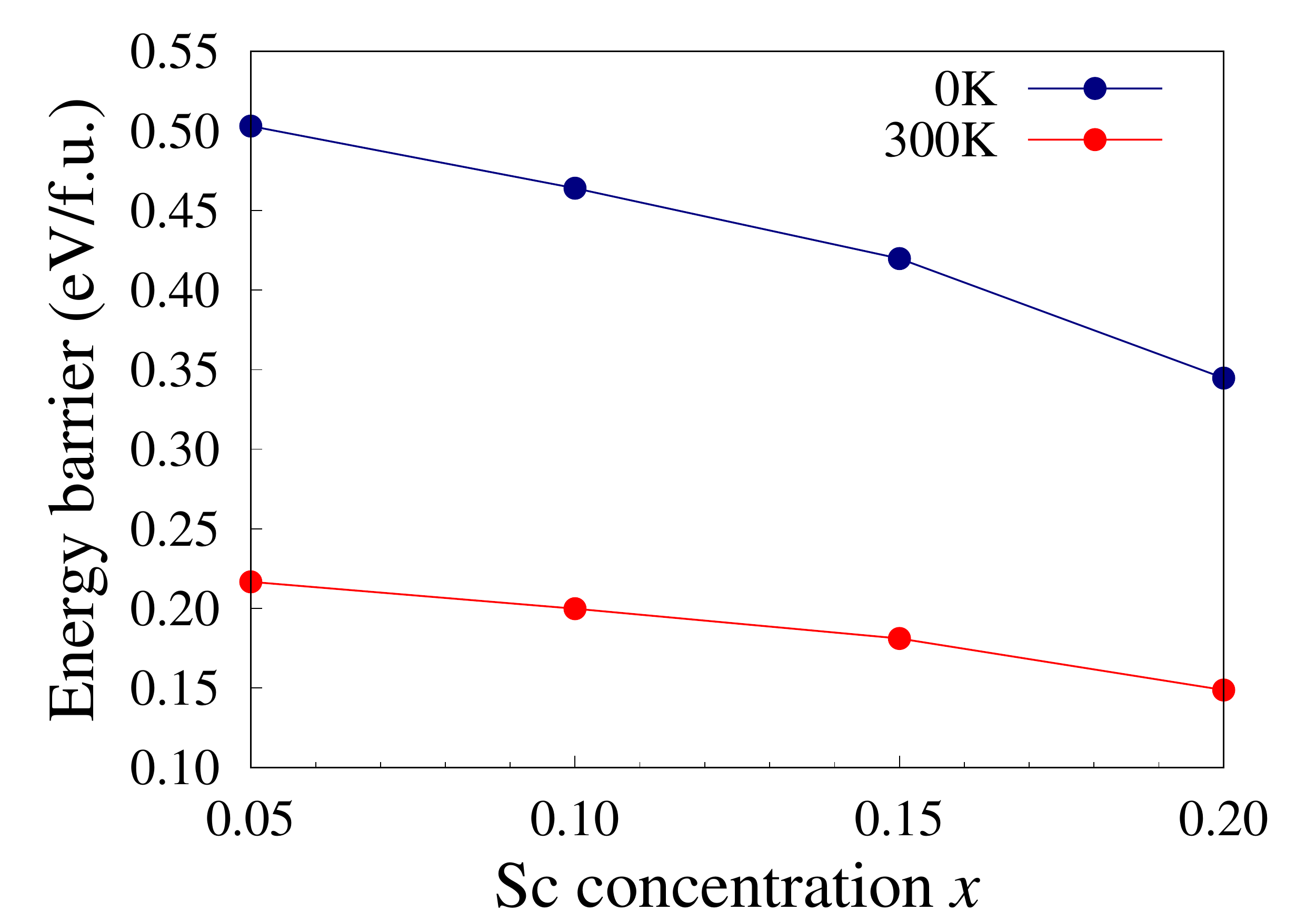}
    \caption{
  Ferroelectric switching barrier (eV/f.u.) for Al$_{1-x}$Sc$_{x}$N at $T$=0 K and at 300 K as a function of Sc concentration.
    }
    \label{T-barriers}
\end{figure}

We use Eq.~(\ref{landau}) to fit our calculated first-principles energies at $T$=0 as a function of polarization, allowing us to extract values for $a(0)$ and $b(0)$ at each Sc concentration $x$.
The curves shown in Fig.~\ref{energy barriers}(a) are a result of these fits.
The energy barrier between the polarized state and the zero-polarization reference state decreases with increasing scandium concentration; for pure AlN the calculated energy barrier is 0.51 eV/f.u., while at 20\% Sc the barrier is decreased to 0.35 eV/f.u. (see Fig.~\ref{T-barriers}).
As noted above, these values constitute an upper bound because we keep the lattice parameters fixed.
Some insight into the effect of allowing lattice relaxation can be obtained by comparing with the results of Ye \emph{et al.}~\cite{ye2021atomistic}, who employed the solid-state nudge elastic band method (ss-NEB)~\cite{sheppard2012generalized} that allows for cell relaxation.
For pure AlN the ss-NEB result for the switching barrier is 0.23 eV/f.u., about half of our value.
For Al$_{0.8}$Sc$_{0.2}$N, the ss-NEB result for the barrier is 0.12 eV/f.u.
Still, this barrier is much too high to allow switching.

To assess the temperature effect on the switching barrier, we follow the Landau theory and assume the parameter $a(T)$ can be written as
\begin{equation}
    a(T)=a_{0} \frac{T-T_{c}}{T_{c}}
\end{equation}
where $a_0$ is a positive constant and $T_c$ is the paraelectric transition temperature.
In the absence of specific information, we adopt the lower-limit $T_c$ value for Al$_{0.64}$Sc$_{0.36}$N, which is 600$^{\circ}$C (Ref.~\onlinecite{fichtner2019AlScN}), for all considered Al$_{1-x}$Sc$_{x}$N alloys.

Figure~\ref{energy barriers}(b) shows the energy barriers at 300 K as a function of Sc concentration.
Again, we see that the ferroelectric switching barrier of Al$_{1-x}$Sc$_x$N decreases with increasing scandium concentration.
Figure~\ref{energy barriers}(c) shows the energy barriers for a 20\% scandium concentration at different temperatures.
The energy barrier is reduced from 0.35 eV/f.u. at 0 K to 0.15 eV/f.u. at room temperature.
The switching barriers at $T$=300 K are included in Fig.~\ref{T-barriers} and compared with the $T$=0 barriers.

Based on our calculations, we can estimate the switching electric field as follows.
If $\vec{E}$ is the applied electric field necessary for switching, then $\omega = \frac{1}{2}\vec{E}\cdot\vec{P}$ is the electric field energy density\cite{griffiths2013introduction}, which we can obtain from the calculated energy $E_b$ at the switching barrier: $\omega=\frac{E_b}{V_\mathrm{f.u.}}$.
The static field part $\frac{1}{2}\epsilon_0\vec{E}^2$ is ignored.
At room temperature, the polarization $P$ of Al$_{0.8}$Sc$_{0.2}$N is 0.85 C/m$^2$ [Fig.~\ref{energy barriers}(b)],
and the corresponding energy barrier $E_b$ is 0.15 eV/f.u. (Fig.~\ref{T-barriers}).
We thus obtain a switching electric field of 26.5 MV/cm.

Experiments have demonstrated ferroelectric switching in Al$_{1-x}$Sc$_x$N starting at Sc concentrations of $x=27\%$.\cite{fichtner2019AlScN}
For $x=27\%$, the polarization was estimated to be 100 $\mathrm{\mu C/cm^2}$=1 C/m$^2$, and
the switching electric field was estimated to be close to 4.5 MV/cm.
The estimate for the switching electric field based on our calculations is thus almost 6 times larger than the experimental one.
The overestimate is expected, based on our assumption of uniform coherent switching  and the lack of cell relaxation when computing the double-well potential in Fig.~\ref{energy barriers}.

\section{\label{summary}Conclusions}

We studied the properties of Al$_{1-x}$Sc$_x$N alloys with the aim of elucidating the underlying mechanisms of the enhancement in piezoelectricity and the observation of ferroelectricity.
We investigated the internal parameter $u$ and mapped energy profiles as a function of scandium concentration.
By decomposing the piezoelectric constant $e_{33}$ into its contributions, we found that the internal strain contribution dominates, which can be attributed to larger displacements associated with Sc ions.
Our comparison between VCA and supercell results indicated that VCA can produce reliable results for polarization in Al$_{1-x}$Sc$_x$N alloys, offering the prospect of performing large-scale simulations that would otherwise be prohibitively expensive.
Our calculations of polarization as a function of internal displacement allowed us to calculate values for the ferroelectric switching barrier at $T$=0.  We also employed Ginzburg-Landau theory to estimate barrier heights at finite temperatures.  While the barrier heights are overestimated (due to the assumption of uniform coherent switching and the absence of cell relaxation), our results offer insight into the variation of barrier height with Sc concentration.

\begin{acknowledgments}
The authors appreciate fruitful discussions with Xie Zhang, Haidong Lu and Simon Fichtner.
This work was supported by the Air Force Office of Scientific Research under Award No. FA9550-18-1-0237.
Use was made of computational facilities purchased with funds from the National Science Foundation (NSF) (No. CNS-1725797) and administered by the Center for Scientific Computing (CSC). The CSC is supported by the California NanoSystems Institute and the Materials Research Science and Engineering Center (NSF DMR 1720256) at UC Santa Barbara.
This work also used the Extreme Science and Engineering Discovery Environment (XSEDE), which is supported by NSF under Grant No. ACI-1548562.

\end{acknowledgments} 

\section*{Data Availability}
The data that support the findings of this study are available from the corresponding author upon reasonable request.

\bibliography{reference,VASP}

\end{document}